# A hypothesis testing framework for the ratio of means of two negative binomial distributions: classifying the efficacy of anthelmintic treatment against intestinal parasites

Denwood, M.J.[1], Innocent, G.T.[2], Prentice, J.C.[3], Matthews, L.[3], Reid, S.W.J.[4], Pipper, C.B.[5], Levecke, B.[6], Kaplan, R.M.[7], Kotze, A.C.[8], Keiser, J.[9,10], Palmeirim, M.[9,10], & McKendrick, I.J. [2]

[1] Department of Veterinary and Animal Sciences, University of Copenhagen, Denmark
[2] Biomathematics and Statistics Scotland, Edinburgh, UK
[3] Institute of Biodiversity Animal Health and Comparative Medicine, University of Glasgow, UK
[4] Royal Veterinary College, UK
[5] LEO Pharma A/S, Ballerup, Denmark
[6] Department of Virology, Parasitology and Immunology, Faculty of Veterinary Medicine, Ghent University, Belgium
[7] Department of Infectious Diseases, College of Veterinary Medicine, University of Georgia, USA
[8] CSIRO Agriculture and Food, St. Lucia, Brisbane, Queensland, Australia
[9] Swiss Tropical and Public Health Institute, Basel, Switzerland
[10] University of Basel, Switzerland

October 15th 2019

**Note:** This is the self-archived pre-peer reviewed version of an article that is currently under review. Once the submitted article is accepted for publication, this archive will be updated with the DOI and link to the published version. In the meantime, this article can be cited as an arXiv e-print using the format given under 'Export citation' at the page on https://arxiv.org/ for this pre-print.

**Abstract**

Over-dispersed count data typically pose a challenge to analysis using standard statistical methods, particularly when evaluating the efficacy of an intervention through the observed effect on the mean. We outline a novel statistical method for analysing such data, along with a statistically coherent framework within which the observed efficacy is assigned one of four easily interpretable classifications relative to a target efficacy: "adequate", "reduced", "borderline" or "inconclusive". We illustrate our approach by analysing the anthelmintic efficacy of mebendazole using a dataset of egg reduction rates relating to three intestinal parasites from a treatment arm of a randomised controlled trial involving 91 children on Pemba Island, Tanzania. Numerical validation of the type I error rates of the novel method indicate that it performs as well as the best existing computationally-simple method, but with the additional advantage of providing valid inference in the case of an observed efficacy of 100%. The framework and statistical analysis method presented also allow the required sample size of a prospective study to be determined via simulation. Both the framework and method presented have high potential utility within medical parasitology, as well as other fields where over-dispersed count datasets are commonplace. In order to facilitate the use of these methods within the wider medical community, user interfaces for both study planning and analysis of existing datasets are freely provided along with our open-source code via: http://www.fecrt.com/framework

**1. Introduction**

Count distributions arise in a number of fields relevant to medicine, and typically present different challenges for analysis than those encountered with continuous distributions, particularly where the mean count is close to zero and the observations are over-dispersed compared to the Poisson distribution. One such challenge arises when assessing the effect of interventions for countable phenomena such as criminal activities [1,2], migraines [3], falls in elderly patients [4], seizures [5,6], MRI lesion counts [7] and intestinal parasites [8–10], where the latter are quantified using counts of parasite eggs in stool. We define the problem first in general terms, and then with a motivating example.

Let $\mu_1$ and $\mu_2$ denote the mean of two over-dispersed count datasets representing the comparator and treatment group, respectively. The ratio of $\mu_2$ to $\mu_1$ is the parameter of interest when measuring the effect of an intervention evaluated against either an alternative treatment or an untreated control. We consider two common study designs: those comparing paired sets of observations from the same individuals before and after treatment (*paired data*), and those comparing a single set of observations from individuals assigned to different treatment/control groups (*unpaired data*). In both of these cases, it is useful to compare $r = 1 - \mu_2 / \mu_1$ to a pre-specified target, but there is currently no clearly-justified statistical framework within which summary statistics and confidence intervals can be interpreted, except in the simplest case when testing that $\mu_1 \neq \mu_2$ [7,11,12]. We note that our focus is fundamentally different from that of the standard hypothesis of equality; we aim to investigate simultaneously two related but distinct questions:

1) Is *r* low enough to conclude that the intervention demonstrates evidenced inadequacy?
2) Is *r* high enough to conclude that the intervention demonstrates evidenced adequacy?

Our motivating example is the analysis of egg reduction rate data (ERR; also known as faecal egg count reduction or FECR) in order to evaluate the efficacy of an anthelmintic treatment against intestinal parasites. Such data is most frequently collected in a *paired* experimental design, but *unpaired* studies comparing randomised treatment/control groups are possible, and a combination of the two situations, whereby multiple treatment arms result in multiple *paired* datasets (one for each treatment arm), are also encountered [e.g. 13,14]. In the *paired* study design, pre-treatment stool samples are obtained from individuals enrolled in the study and the number of parasite eggs within a fixed volume of stool is enumerated. A number of days after the administration of an anthelmintic treatment, a post-treatment stool sample is obtained from each of the individuals, and the number of parasite eggs within a fixed quantity of stool is again enumerated [15]. One statistic of interest is the ERR, which is defined as one minus the ratio of arithmetic means between the pre-treatment data ($\mu_1$) and post-treatment data ($\mu_2$), thus reflecting the average efficacy of the drug in these individuals. An alternative statistic of interest is the cure rate, although we consider only the ERR here because the cure rate is arguably less appropriate for assessing drug efficacy [9,16]. Typically, the goal of quantifying the ERR is either to assess the efficacy of novel anthelminthic treatments or regimens [13], or to provide evidence for reduced efficacy compared to the published expected efficacy of the same drug in a naïve population, i.e. to determine if either anthelmintic resistance is present or the drugs being used are of poor quality [9,10,17]. Although the World Health Organisation guidelines [18] only discuss the use of point estimates, the standard approach to analysis within the medical literature is to generate 95% confidence intervals for the drug efficacy using a non-parametric bootstrap [13,19]. In veterinary parasitology, a variety of parametric approximation methods are used, such as that given by Coles et al. [20] and later expanded by Pepper et al. [21], the method of Levecke et al. [10], and the method of Dobson et al. [22]. However, only one of these methods can be used when the observed efficacy equals 100% due to the post-treatment mean and variance of zero.

Alternative methods have been suggested based on Markov chain Monte Carlo (MCMC) [23–26], but the correct use of such computationally intensive methods requires statistical and computational skills and training that may be out of scope for the typical practitioner [27].

In addition to the lack of a standardised method to calculate 95% confidence intervals, there is also a current lack of consensus regarding the interpretation of results based on these confidence intervals. In a veterinary parasitology conference abstract, Pepper et al. [21] used a simulation study to explore different interpretations and concluded that there were a number of inadequacies in standard approaches, arising from the failure to quantify and incorporate uncertainty. They recommended using a test of inferiority, i.e. a one-sided test of *r* compared to a pre-specified target value. They also noted the importance of powering studies appropriately, given what they perceived as a low probability of detecting situations where the true efficacy was below the target. Accordingly, the same authors used a framework based on an inferiority test to interpret a simulation study [28]. However, these studies are based on the empirical performance of different classification methods rather than a rigorous justification of the underlying statistical issues. To the authors' knowledge, no such justification currently exists in either the medical or veterinary parasitology literature, or elsewhere within related fields.

Our objective is therefore to outline and justify a statistical framework to evaluate the ratio of means between two data series drawn from negative binomial distributions. We implement this framework along with three currently available statistical tests, as well as two novel statistical methods, and compare the statistical properties of these methods by use of simulation. We illustrate our approach by analysing ERR data from one arm of a randomised controlled trial representing the efficacy of the anthelmintic mebendazole against three intestinal parasites in children; hookworm, *Ascaris lumbricoides* and *Trichuris trichiura*.

## 2. Data

Our analytical methods are illustrated using a dataset obtained from a randomised controlled trial in children on Pemba Island, Tanzania, which is registered at ClinicalTrials.gov (NCT03245398) and has previously been described by Palmeirim et al. [14]. The purpose of the original study was to compare single-dose and multiple-dose mebendazole treatment, but for the purposes of this manuscript, we only use data from the 91 individuals who completed the study according to the protocol for single-dose mebendazole treatment. Two stool samples were taken from each child on different days, and were examined by preparing and screening duplicate Kato-Katz thick smears [29] for each stool sample (each Kato-Katz thick smears represents approximately 41.7 mg of stool). The slides were examined under a light microscope and the total count over the four slides for each child was then recorded for the eggs of three parasites (hookworm, *A. lumbricoides* and *T. trichiura)*. All 91 children were positive for hookworm at pre-treatment, 46 children were positive for *A. lumbricoides*, and 46 children were positive for *T. trichiura*. Each of the children was treated with the anthelmintic mebendazole before another two consecutive stool samples per subject were taken 14-21 days post-treatment and analysed as for the pre-treatment samples. This resulted in a total of 546 count observations from 91 individuals (two samples each relating to three parasites). Although Palmeirim et al. [14] reported data in terms of eggs per gram for each parasite (after multiplying observed counts by a constant of six), we have preserved the raw count observations for the purposes of this manuscript, and present mean values accordingly. Appropriate ERR targets for mebendazole for use with these parasites are given by the World Health Organisation [30; table A2.2] and Levecke et al. [10] as: 70% for hookworm, 95% for *A. lumbricoides* and 50% for *T. trichiura*. The objective of this exercise was to evaluate the observed efficacy of the drug by comparing the observed ERR to these published target figures, so that the evidence for reduced anthelmintic efficacy in this population can be assessed.

### 3. Statistical framework

We propose a formal statistical framework that can be used to interpret the observed counts $x_i$ and $y_j$ in terms of the quantity of interest $r = 1 - \mu_2 / \mu_1$. Our example focuses on application to *paired data*, but we note that the framework also holds for use with *unpaired data*, either with treatment *vs*. control groups or with direct comparison between different treatment groups if the target efficacy is interpreted as the ratio of post-treatment mean counts between two treatment arms. If the purpose of the exercise is simply to establish estimates for *r*, then a straightforward point estimate and confidence interval associated with this estimator will suffice. However, we anticipate a further desire to classify the biological implications of the result according to a standardised interpretative framework illustrated by the two questions given in the introduction:

1) Is *r* low enough to conclude that the intervention demonstrates evidenced inadequacy?
2) Is *r* high enough to conclude that the intervention demonstrates evidenced adequacy?

We follow Pepper et al. [21] in suggesting that a one-sided inferiority test formulation is appropriate to address question (1), i.e. to assess whether there is statistically significant evidence that the observed mean efficacy is lower than desired. The most relevant location parameter against which to compare *r* for this purpose is a threshold $T_I$ equal to the expected efficacy *E*, e.g. 0.95 for *A. lumbricoides*.

A test of superiority of the observed mean efficacy relative to an appropriate location parameter can be formulated in order to address question (2), although the authors are not aware of any previous study in which this has been done. In any case, we believe that a non-inferiority test is conceptually better justified. Such tests form part of the regulatory regime for producing generic pharmaceutical products, where it is necessary to show that the generics will not be appreciably worse than the original products [31–34]. We believe that there is a clear parallel between this situation and the desire to demonstrate that the efficacy of an anthelmintic compound observed 'in the field' is not appreciably worse than the expected efficacy of the same drug. Such a hypothesis requires the definition of an additional quantity: the margin of equivalence $\delta$, which quantifies the idea of 'not appreciably worse' as used above. We note that the margin of equivalence in a non-inferiority test can be made arbitrarily small; if the margin is set to zero, then the properties of the non-inferiority test will be identical to those of a test of superiority. The margin of equivalence $\delta$ can usefully be defined as the difference between the target efficacy *E* and the maximum value for *r* that would be tolerated as consistent with an effective intervention. For example, if *E* is taken as 0.95 for *A. lumbricoides* and a non-inferiority threshold $T_A$ is defined as a minimum acceptable efficacy of 0.90, then $\delta = 0.05$. In a situation where two drugs are to be compared directly by randomisation of individuals between the drugs, then the natural choice for the target *E* is a value of 0, reflecting an identical efficacy between drugs (so that the expected ratio $\mu_2 / \mu_1$ is equal to 1), and the choice of $\delta$ follows the standard procedure for equivalence testing [34].

Based on these two tests, and assuming that interpretation of the results will depend on the positioning of lower and upper confidence interval estimates for *r* obtained via some valid statistical method, it is straightforward and instructive to consider the different possible typologies that might arise in practice (Figure 1). There are ten different typologies in total, although these can be grouped into four typology groups based solely on the results of the two tests previously defined. It is important to note that previous interpretative approaches within the veterinary parasitology literature [20,25,28,35] have used the point estimate as part of the classification system (which would require us to provide an interpretation of each of the ten typologies), but we reject this strategy because it mixes two very different types of information relating to the population and the sample: the confidence interval, which quantifies uncertainty about the true underlying efficacy (population information), and the point estimate, which is simply a summary of what was observed in the study (sample information). For typologies 1a, 1b & 1c, the upper confidence bound for *r* is below the target $T_I$, which is sufficient to conclude that the observed efficacy is statistically significantly inadequate, as defined previously. For typologies 4a, 4b & 4c, the lower confidence bound is above the target $T_A$ so we are able to conclude that *r* is statistically significantly adequate; typology 4c reflects the more extreme situation where *r* would be statistically significantly greater than the published naïve efficacy *E*, which may be unlikely to occur in practice. This illustrates the rationale behind using a non-inferiority hypothesis test within our framework; if a test of superiority were used instead, typology 4c would be the only situation in which the efficacy would be assessed as adequate. Typologies 2a, 2b & 2c reflect situations where the confidence interval for *r* is so wide that there is no evidence that the true value can be excluded from any of the parameter regions – a potential consequence of an inadequate sample size. Finally, typology 3 reflects a situation where there is statistically significant evidence that *r* lies between $T_A$ and $T_I$, which can only occur when there is a sufficiently large sample size that the entire confidence interval is within these limits. The authors believe that it is appropriate for this typology to have a clear and distinct designation since, should it arise in practice, this would reflect a very specific situation: where there is statistically significant evidence that the treatment has less efficacy than the target, but where it is also within the specified margin of equivalence. In our motivating example, this would correspond to a biological interpretation of early stage and/or low-level reduction in drug efficacy due to e.g. emerging anthelmintic resistance or sub-optimal drug quality.

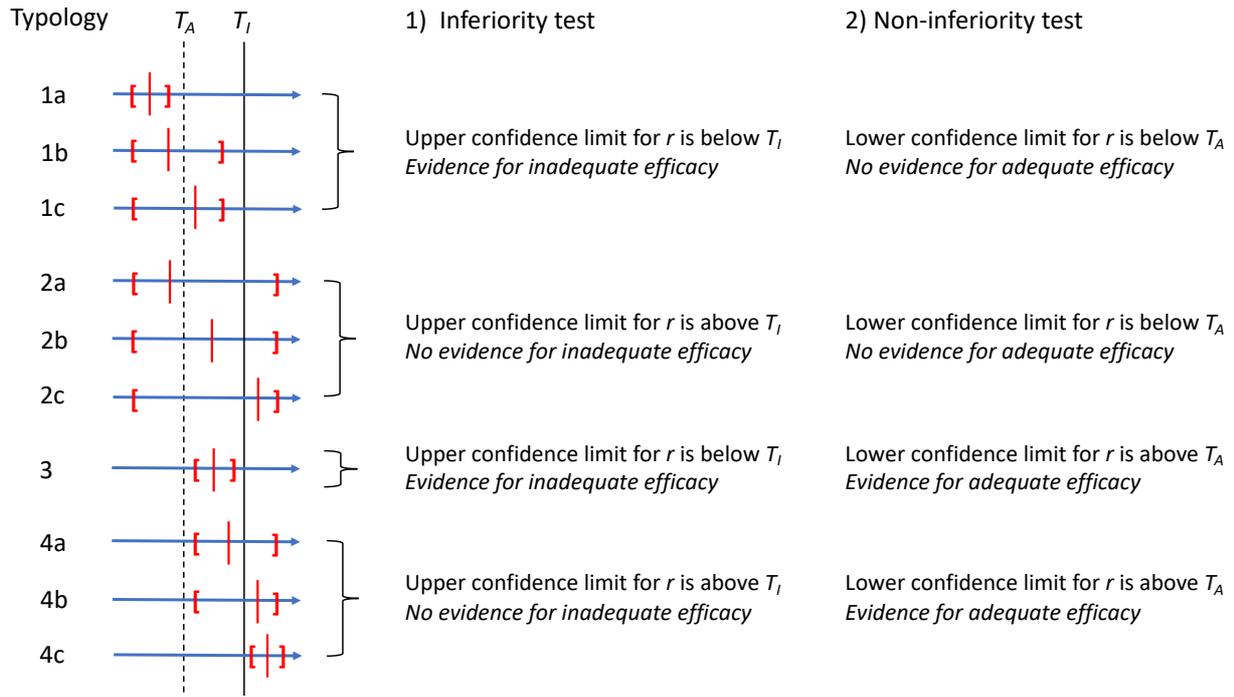

Figure 1: Possible typologies that result from assessing the position of the point estimate $\hat{r}$ (red bar) and confidence intervals for *r* (red brackets), relative to the two thresholds $T_A$ (dashed line) and $T_I$ (solid line). Equivalent interpretations within typology groups are indicated using right braces.

We also note that it is not necessary to know the precise estimates of the confidence intervals to classify to the four topologies; only the binary results of the two comparisons of lower CL $\geq T_A$ and upper CL $< T_I$. We can therefore define the classifications via one-sided hypothesis tests, as follows:

1) Inferiority test: there is evidence for reduced efficacy ($r < E$) if:
    a. Upper 95% CL $< T_I$
   
   Or, equivalently:
   
    b. We can reject the null hypothesis that $r \geq T_I$ with $p < 0.025$

2) Non-inferiority test: there is evidence for adequate efficacy ($r \geq E - \delta$) if:
    a. Lower 95% CL $\geq T_A$
   
   Or, equivalently:
   
    a. We can reject the null hypothesis that $r < T_A$ with $p < 0.025$

Where only one of these two criteria are met, the result should be used to categorise the result as either **adequate efficacy** (typology 4) or **reduced efficacy** (typology 1), respectively. These classifications are based on positive evidence of either acceptable or reduced efficacy. Where neither criterion is met, the result should be categorised as **inconclusive** (typology 2) on the basis that there is insufficient evidence to conclude either that $r \geq T_A$ or that $r < T_I$. We follow Torgersen et al. [24] and Geurden et al. [25] in the use of this inconclusive categorisation, which we believe is valuable given that sufficient evidence for the positive classifications defined above is not always available in practice. Depending on the sample size and chosen value of $\delta$, it is also possible for both criteria to be met: a further category, **borderline efficacy** (typology 3), is therefore defined to describe situations where there is statistical evidence that $r$ is below the threshold $T_I$ but also above the threshold $T_A$.

*Hypothesis testing for the ratio of negative binomial distributions*## 4. Statistical methods

*4.1. Notation and general remarks*

The following terms are used throughout this section:

- *efficacy*: In this manuscript the term *efficacy* is simply used as a synonym for the arithmetic mean egg reduction rate (ERR), although we note that the parasitological definition of anthelmintic efficacy is a more complex issue.
- $x_i$: A set of pre-treatment count observations from individuals $i$ in $1 \ldots N_1$, distributed according to a negative binomial distribution with mean $\mu_1$ and shape parameter $k_1$.
- $y_j$: A set of post-treatment count observations from individuals $j$ in $1 \ldots N_2$, distributed according to a negative binomial distribution with mean $\mu_2$ and shape parameter $k_2$.
- *paired data:* A study design where if $i = j$, the observations $x_i$ and $y_j$ will have been collected from the same individual. In this case, we assume that $N_1 = N_2$.
- *unpaired data:* A study design where an observation of any individual $i$ is assumed to be independent of subsequent observations of all individuals $j$. In this case, we do not necessarily assume that $N_1 = N_2$.
- $r = 1 - \mu_2 / \mu_1$: The quantity of interest, representing the efficacy (ERR) of the treatment.
- $\hat{r} = 1 - \frac{N_1 \sum_{j=1}^{N_2} y_j}{N_2 \sum_{i=1}^{N_1} x_i} = 1 - \frac{\bar{y}}{\bar{x}}$: The maximum likelihood estimator of $r$, which corresponds to the standard calculation of the arithmetic mean ERR.
- $\widehat{\mu_1} = \frac{1}{N_1} \sum_{i=1}^{N_1} x_i = \bar{x}$: The maximum likelihood estimator of $\mu_1$ (and equivalently $\mu_2$).
- $E$: The target for $r$ based on the expected effect of the intervention in optimal circumstances, which could be interpreted as full susceptibility to the anthelmintic compound, e.g. a target efficacy of 95% for *A. lumbricoides* would correspond to the expectation that $\mu_2 \leq (1-0.95)\mu_1$ under a fully effective treatment.
- $\delta$: A non-inferiority margin below $E$ indicating the minimum value of $r$ that would be considered to be consistent with a fully efficacious intervention.
- $T_I = E$: The inferiority test threshold.
- $T_A = E - \delta$: The non-inferiority test threshold.

A number of methods have previously been proposed to generate 95% confidence intervals for the ratio of means between two count distributions: we briefly outline these below for the case of *paired data*, therefore assuming that $N = N_1 = N_2$. For simplicity, we will also assume that a single pre-treatment count and a single post-treatment count are available for each individual, but we note that each method can also be used after summing multiple pre-treatment and/or post-treatment observations for each individual. For example, where each $x_i$ consists of $M_1$ replicate count observations within individual $i$ (e.g. duplicate or quadruplicate Katz Katz thick smears), then summing these replicate observations within each $x_i$ results in a Negative Binomial distribution with mean equal to $M_1 \cdot \mu_1$ and an over-dispersion parameter equal to $M_1 \cdot k_1$ [36]. If each $y_j$ consists of $M_2$ replicate count observations within individual $j$, then these can be summed in a similar manner. The analysis can then be performed on the sums of these replicate observations, conditional on the number of replicates $M_1$ being consistent across individuals $i$, and the number of replicates $M_2$ being consistent across individuals $j$.

*4.2. Existing computationally simple methods*

A normal approximation method is given in the appendix of Coles et al. [20] and later justified by Pepper et al. [21]. Using our notation, the upper and lower 95% confidence limits for $r$ are given by $100 \cdot \left(1 - \frac{\mu_2}{\mu_1}\right) e^{\pm t\sqrt{V}}$ where $V = \frac{\sigma_1^2}{N\mu_1} + \frac{\sigma_2^2}{N\mu_2} - \frac{2 \cdot cov_{1,2}}{N\mu_1\mu_2}$ and $t$ is the 97.5 percentile point for Student's t distribution with degrees of freedom equal to $N - 1$. The confidence intervals are indeterminate when $\sum y_j = 0$, and previous work has identified that egg count data in animals are not well approximated by a normal distribution [37], although the method has empirically been shown to work well within a relatively narrow range of parameter values when $\sum y_j > 0$ [23]. We refer to this as the World Association for the Advancement of Veterinary Parasitology or **WAAVP** method.

An alternative approximation based on the gamma distribution with parameters obtained using a delta method approximation is given by Levecke et al. [10]. The upper and lower 95% confidence limits for $r$ are given by the 2.5th and 97.5th quantiles of the gamma distribution with parameters $\alpha = \frac{(1-\hat{r})^2}{V}$ and $\beta = \frac{V}{1-\hat{r}}$, where $V = \frac{1}{N}\left(\frac{\mu_2}{\mu_1}\right)^2 \left(\frac{\sigma_1^2}{\mu_1^2} + \frac{\sigma_2^2}{\mu_2^2} - 2 \cdot \frac{\rho\sqrt{\sigma_1^2\sigma_2^2}}{\mu_1\mu_2}\right)$ and $\rho$ represents the within-individual correlation between pre- and post-treatment samples. We refer to this as the **Gamma** method.

An exact Bayesian method based on conjugate priors has been proposed by Dobson et al. [22] based on the method of estimating confidence intervals for a binomial proportion given by Brown et al. [38]. Assuming that $y_i \sim$ Binomial($x_i$, *1-r*) then the beta distribution can be used as a conjugate prior to fully describe the posterior as:

$$(1-r) \sim \text{Beta}\left(\alpha_0 + \sum_{j=1}^{N} y_j, \quad \beta_0 + \left(\sum_{i=1}^{N} x_i\right) - \left(\sum_{j=1}^{N} y_j\right)\right)$$

This method requires definition of the parameters $\alpha_0$ and $\beta_0$ in the prior, for which Dobson et al. [22] justify the use of values of 1 based on the empirical remark that it generates more conservative lower confidence intervals than the Jeffrey's priors of 0.5 suggested by Brown et al. [38]. This method ignores the correlation between paired $x_i$ and $y_i$, and also assumes that the number of 'trials' is a fixed quantity, i.e. it does not model the potentially appreciable uncertainty in $\sum x_i$. As a consequence, the confidence intervals generated are generally too narrow. Dobson et al. [22] note this based on empirical observations, and advocate the use of 99% confidence intervals in place of 95% confidence intervals to compensate for the effect. However, this method does not require that $\sum y_j > 0$, which is an important potential advantage. We refer to this as the **Binomial** method.

### 4.3. Asymptotic distribution of $\hat{r}$

We note that the means of $x$ and $y$ will be asymptotically distributed as:

$$\begin{bmatrix} \bar{x} \\ \bar{y} \end{bmatrix} \approx \text{Normal}\left(\begin{bmatrix} \mu_1 \\ (1-r)\mu_1 \end{bmatrix}, \frac{1}{N}\mathbf{S}\right)$$

Where:

$$\mathbf{S} = \begin{bmatrix} \mu_1 + \frac{\mu_1^2}{k_1} & 0 \\ 0 & \mu_1 - r\mu_1 + \frac{(\mu_1 - r\mu_1)^2}{k_2} \end{bmatrix}$$

Then, using the delta method [39], the maximum likelihood estimator for *1- r* is asymptotically distributed as:

$$1 - \hat{r} \approx \text{Normal}\left(1 - r, \frac{1}{N}\Delta^2\right)$$

where,

$$\Delta^2 = \begin{bmatrix} \frac{1-r}{\mu_1} & \frac{1}{\mu_1} \end{bmatrix} \mathbf{S} \begin{bmatrix} (1-r)/\mu_1 \\ 1/\mu_1 \end{bmatrix}$$

The matrix **S** can be estimated as:

$$\hat{\mathbf{S}} = \begin{bmatrix} \frac{1}{N-1}\sum_{i=1}^{N}(x_i - \bar{x})^2 & 0 \\ 0 & \frac{1}{N-1}\sum_{i=1}^{N}(y_i - \bar{y})^2 \end{bmatrix}$$

and using maximum likelihood estimators for $1 - \hat{r} = \bar{y}/\bar{x}$ and $\mu_1 = \bar{x}$, we obtain:

$$\widehat{\Delta^2} = \begin{bmatrix} \frac{1-\hat{r}}{\bar{x}} & \frac{1}{\bar{x}} \end{bmatrix} \hat{S} \begin{bmatrix} (1-\hat{r})/\bar{x} \\ 1/\bar{x} \end{bmatrix} = \left(\frac{\bar{y}}{\bar{x}^2}\right)^2 \frac{1}{N-1} \sum_{i=1}^{N}(x_i - \bar{x})^2 + \left(\frac{1}{\bar{x}}\right)^2 \frac{1}{N-1} \sum_{i=1}^{N}(y_i - \bar{y})^2$$

For *paired data,* it is also necessary to take account of the correlation between pre- and post-treatment samples from the same individual. We do this by scaling the variance estimates $\frac{1}{N-1}\sum_{i=1}^{N}(x_i - \bar{x})^2$ and $\frac{1}{N-1}\sum_{i=1}^{N}(y_i - \bar{y})^2$ by 1 minus the correlation between paired $x_i$ and $y_i$. This allows for the calculation of asymptotically unbiased and efficient confidence intervals for *r*, but the estimates may be poor for small sample sizes. We refer to this as the **Asymptotic** method.

*4.4. Exact distribution of an alternative test statistic*

Although *r* is typically the parameter of interest, the hypothesis testing framework allows us to consider any related test statistic for which the expected distribution under the null hypotheses can be defined. As such, we note that $\sum_{j=1}^{N_2}(y_j - \bar{y})^2 = 1 - r\left(\sum_{i=1}^{N_1}(x_i - \bar{x})^2\right)$ is a potentially valid test statistic, and that:

$$\sum_{j=1}^{N_2} y_j \sim \text{NegBin}(N_2 k_2, p_2)$$

where:

$$p_2 = \frac{k_2}{k_2 + (1-r)\mu_1}$$

The quantities $N_2$ and $r$ are fixed and defined by the null hypotheses, respectively. The quantities $k_2$ and $\mu_1 = k_1 \frac{1-p_1}{p_1}$ are unknown but can be estimated from the data. Following a similar rationale to that of Dobson et al. [22], we note that the Bayesian posterior distribution of $p_1$ can be defined using a beta conjugate prior:

$$p_1 \sim \text{Beta}\left(\alpha_0 + \sum_{i=1}^{N_1} x_i, \beta_0 + k_1 N_1\right)$$

Again, $N_1$ is known, and the prior quantities $\alpha_0$ and $\beta_0$ can be set to take small values, reflecting a minimally informative prior belief for the parameter $p_1$. By taking an indefinite integral of these two distributions, the expected distribution of the test statistic can be approximated using a beta negative binomial distribution:

$$\sum_{j=1}^{N_2} y_j \approx \text{BetaNegBin}(\alpha_2, \beta_2, N_2 k_2)$$

Where $\alpha_2$ and $\beta_2$ can be calculated as non-linear functions of $r$, $\alpha_0$, $\beta_0$, $\sum_{i=1}^{N_1} x_i$, $N_1$, $k_1$ and $k_2$ (see the statistical appendix for further details). For *paired data,* it is necessary to take account of the correlation between pre- and post-treatment samples from the same individual. We do this by scaling the over-dispersion parameters $k_1$ and $k_2$ by 1 minus the correlation between paired $x_i$ and $y_i$. This method accounts for uncertainty in $\mu_1$, but ignores uncertainty in $k_1$ and $k_2$, which are simply taken as point estimates from the data (we calculate these by maximum likelihood). The impact of assuming that $k_1$ and $k_2$ are fixed is assessed by simulation in section 6. Where $k_2$ cannot be estimated from the data (i.e. when $\sum_{j=1}^{N_2} y_j = 0$), then the additional simplifying assumption of $k_2 = k_1$ can be made, which permits the use of this method in the important case that $\sum y_j = 0$. We refer to this as the Beta Negative Binomial or **BNB** method.

Unlike the other methods described above, the **BNB** method cannot be used to generate confidence intervals for *r*, but it can be used as the basis for hypothesis testing within the framework specified in section 3, given values of *r* against which to test the observed data. Specifically, the observed test statistic $\sum_{j=1}^{N_2} y_j$ can be compared to the expected distribution of the test statistic under the null hypotheses of:

1) One-sided inferiority test of evidence for reduced efficacy (*r* < *E*)

    H<sub>0</sub>:  $r \geq T_I$

    H<sub>1</sub>:  $r < T_I$

    Where the associated p-value $p_I$ is defined as the probability of observing values of the test statistic $\sum y$ that are greater than or equal to the observed $\sum_{j=1}^{N_2} y_j$ under a beta negative binomial distribution with $r = T_I$ and other parameters as derived above.

2) One-sided non-inferiority test of evidence for adequate efficacy ($r \geq E - \delta$)

    H<sub>0</sub>:  $r < T_A$

    H<sub>1</sub>:  $r \geq T_A$

    Where the associated p-value $p_A$ is defined as the probability of observing values of the test statistic $\sum y$ that are less than or equal to the observed $\sum_{j=1}^{N_2} y_j$ under a beta negative binomial distribution with $r = T_A$ and other parameters as derived above.

The classifications specified in section 3 can then be applied to the results of these p-values, with significance thresholds of p < 0.025 used here for equivalence with the use of 95% confidence intervals. This allows for comparison of the **BNB** method to the **WAAVP**, **Gamma**, **Binomial** and **Asymptotic** methods.

## 5. Application

*5.1. Analysis methods*

Data derived from the three paired datasets described in section 2 were analysed using the five analytical methods described in section 4 (**BNB**, **WAAVP**, **Gamma**, **Binomial**, and **Asymptotic**). Mebendazole efficacy targets of 70% for hookworm, 95% for *A. lumbricoides* and 50% for *T. trichiura* were used, as recommended by the World Health Organisation [30]. There is currently no standard recommendation for an appropriate non-inferiority margin for use with this anthelmintic, so we used an arbitrary value of 0.05 for consistency with efficacy threshold values previously recommended in the veterinary literature [20]. However, a value of e.g. 0.1 is equally justifiable for situations where this would reflect the maximum reduction in efficacy that is clinically acceptable [34]. Depending on the analytical method, either p-values (for the BNB method) or 95% confidence intervals (for other methods) were used to generate a classification for the observed efficacy as previously described.

*5.2. Analysis results*

Estimated efficacies (defined here as arithmetic mean ERR) of 53%, 100% and 49% were calculated for hookworm, *A. lumbricoides* and *T. trichiura,* respectively. Estimates of the over-dispersion parameters $k_1$ and $k_2$ were generated for each dataset using maximum likelihood estimation. These estimates are given along with the pre- and post-treatment mean egg counts and the correlation between pre- and post-treatment counts in Table 1.

Table 1: Estimates of pre-treatment arithmetic mean, post-treatment arithmetic mean, pre- and post-treatment over-dispersion parameters $k_1$ and $k_2$ and the correlation between pre- and post-treatment data for three intestinal parasites based on samples from 91 children from Pemba Island (Tanzania). These numbers are presented in terms of the raw total egg counts rather than eggs per gram as given by Palmeirim et al. [14].

|  | Hookworm | *Ascaris lumbricoides* | *Trichuris trichiura* |
| --- | --- | --- | --- |
| Pre-treatment mean | 74 | 1,255 | 162 |
| Post-treatment mean | 35 | 0 | 83 |
| $k_1$ | 0.84 | 0.08 | 0.92 |
| $k_2$ | 0.58 | -- | 0.53 |
| Correlation | 0.65 | -- | 0.68 |

Results from the five analytical methods applied to the three datasets are shown in Table 2. All methods yielded a classification of reduced efficacy (typology 1) for hookworm: the observed efficacy of 53% being significantly less than the target value of 70%. Confidence intervals were also qualitatively similar between the methods producing 95% confidence intervals, with the exception of the Binomial method, which generates confidence intervals that are spuriously narrow [35]. For *A. lumbricoides*, the WAAVP, Gamma and Asymptotic methods failed to calculate uncertainty due to the 100% observed reduction. The remaining BNB and Binomial methods both classified the efficacy as adequate (typology 4). For *T. trichiura*, the inconclusive classification was assigned using all but one of the methods: the observed efficacy of 49% being neither significantly lower than the target value of 50% nor significantly within the margin of equivalence (typology 2). The only exception was the Binomial method, which classified the efficacy as adequate, again reflecting that this approach overstates the statistical confidence.

Table 2: 95% confidence limits (or p-values) and associated classifications produced by five methods of statistical analysis for the egg count reduction of three intestinal parasites based on samples from 91 children from Pemba Island (Tanzania).

| Statistical Method | Hookworm [Observed: 53%] [Target: 70%] | | *Ascaris lumbricoides* [Observed: 100%] [Target: 95%] | | *Trichuris trichiura* [Observed: 49%] [Target: 50%] | |
|---|---|---|---|---|---|---|
| | Estimate | Classification | Estimate | Classification | Estimate | Classification |
| WAAVP | LCL: 39% UCL: 63% | Reduced | LCL: -- UCL: -- | NA: $\sum y_j = 0$ | LCL: 32% UCL: 61% | Inconclusive |
| Gamma | LCL: 40% UCL: 64% | Reduced | LCL: -- UCL: -- | NA: $\sum y_j = 0$ | LCL: 34% UCL: 62% | Inconclusive |
| Binomial | LCL: 51% UCL: 54% | Reduced | LCL: 99% UCL: 100% | Adequate | LCL: 48% UCL: 50% | Adequate |
| Asymptotic | LCL: 44% UCL: 62% | Reduced | LCL: -- UCL: -- | NA: $\sum y_j = 0$ | LCL: 42% UCL: 55% | Inconclusive |
| BNB | $p_A=1.000$ $p_I<0.001$ | Reduced | $p_A<0.001$ $p_I=1.000$ | Adequate | $p_A=0.261$ $p_I=0.394$ | Inconclusive |

## 6. Numerical validation

*6.1. Validation Methods*

The five different statistical methods outlined in section 4 vary in terms of limitations and assumptions, but are sufficiently computationally simple that their relative performance can be assessed using a simulation study. It is therefore useful to undertake a short numerical validation of each of these methods with parameter values similar to those encountered in the example datasets in order to ascertain the type I and type II error rates for the inferiority and non-inferiority tests associated with each method.

Over-dispersed count data were simulated from negative binomial distributions corresponding to the two pre- and post-treatment datasets for each of the three parasites, with simulation parameter values for pre-treatment mean count, $k_1$, $k_2$ and correlation between pre- and post-treatment counts as given in Table 1. The correlation was simulated by drawing bivariate gamma values using the method described by Nadarajah and Gupta [40], and subsequently simulating the observed counts using a Poisson variate based on these. It was not possible to calculate $k_2$ or the correlation between paired $x_i$ and $y_i$ for *A. lumbricoides* due to the 100% observed reduction, so values of $k_2 = 0.64\ k_1$ and correlation = 0.67 were assumed based on estimates of $k_2 / k_1$ and correlation obtained from the other two species. A total of 10,000 replicate pre-treatment datasets were simulated for each parasite, each with corresponding post-treatment datasets using 1,001 different values of $r$ in the set {0.000, 0.001, 0.002, … 0.998, 0.999, 1.000} to represent the full spectrum of possible comparative efficacy values. This exercise was repeated with sample sizes of N = 91 (to match the example dataset), as well as N = 20 and N = 1,000 for comparison. Each of these ~90 million datasets were then analysed with each of the five statistical methods. The frequencies of each individual hypothesis test result were retained along with classification frequencies for each combination of analysis method, simulated parasite parameters, $r$ and sample size. Where a simulated reduction of 100% was obtained, neither hypothesis was assumed to be rejected for the WAAVP, Gamma and Asymptotic methods. The code required to run these simulations, including both data simulation and statistical analysis methods, was implemented in C++ using an Rcpp interface [41] within the "bayescount" package for R [42].

*6.2. Validation Results*

Generation and analysis of the datasets took approximately 160 μs per N = 20 dataset, 660 μs per N = 91 dataset and 7 ms per N = 1,000 dataset on a five-year-old 3.5GHz Xeon-class workstation. Profiling of the relevant C++ code involved indicated that the majority of the time was used in simulating the datasets and re-estimating the over-dispersion parameters via maximum likelihood.

The most relevant type I error rates, given our assumed sample size of 91 at simulated efficacy values corresponding to $T_I$ (the infimum of the set $\{r \in \mathbb{R} \mid E \geq r \geq 1\}$) for the inferiority test and $T_A$ (the supremum of the set $\{r \in \mathbb{R} \mid 0 \geq r > E - \delta\}$) for the non-inferiority test, are given in Table 3. The notional 2.5% type I error rates were generally very similar across the BNB, WAAVP and Gamma methods, although the error rate of the Gamma method exceeded that of the other methods for the non-inferiority test with *A. lumbricoides*, and the error rate of the BNB method exceeded that of the other methods for the inferiority test with *T. trichiura*. However, the relatively high error rate associated with the inferiority test for *A. lumbricoides* is of particular concern for the WAAVP (9.7%), Gamma (8.1%) and BNB (9.4%) methods. The type I error rates of the Asymptotic and Binomial methods far exceeded the notional rate of 2.5% for all parasite/test combinations, up to a maximum of 46.9% for the Binomial method and 17% for the Asymptotic method. Examples of type II error rates for a sample size of 91 at selected efficacy values are given in Table 4. These efficacy values were chosen to correspond to $T_I$ for the non-inferiority test, reflecting the power to detect non-inferiority at a true efficacy equal to *E,* and the supremum of the set $\{r \in \mathbb{R} \mid 0 \geq r > E - \delta\}$ for the inferiority test. The type II error rates of the BNB, Gamma and WAAVP methods were again similar, although the Gamma and WAAVP methods had lower type II error rates for the non-inferiority test with *A. lumbricoides*.

Table 3: Type I error rates for each of the five statistical methods applied to simulated data with a sample size of N = 91, corresponding to estimated parameter values from egg reduction rates of three intestinal parasite species based on samples from 91 children from Pemba Island (Tanzania). The values of *r* chosen correspond to the infimum of the set $\{r \in \mathbb{R} \mid E \geq r \geq 1\}$) for the inferiority test, and the supremum of the set $\{r \in \mathbb{R} \mid 0 \geq r > E - \delta\}$ for the non-inferiority test.

| Parasite species | Test | r | WAAVP | Gamma | Binomial | Asymptotic | BNB |
|---|---|---|---|---|---|---|---|
| Hookworm | Inferiority | 0.70 | 0.036 | 0.033 | 0.327 | 0.094 | 0.038 |
| | Non-Inferiority | 0.65 | 0.021 | 0.029 | 0.346 | 0.116 | 0.021 |
| *A. lumbricoides* | Inferiority | 0.95 | 0.097 | 0.081 | 0.460 | 0.170 | 0.094 |
| | Non-Inferiority | 0.90 | 0.024 | 0.053 | 0.469 | 0.167 | 0.014 |
| *T. trichiura* | Inferiority | 0.50 | 0.039 | 0.035 | 0.424 | 0.130 | 0.048 |
| | Non-Inferiority | 0.45 | 0.019 | 0.028 | 0.426 | 0.129 | 0.020 |

Table 4: Type II error rates for each of the five statistical methods applied to simulated data with a sample size of N = 91, corresponding to estimated parameter values from egg reduction rates of three intestinal parasite species based on samples from 91 children from Pemba Island (Tanzania). The values of *r* chosen correspond to *E* for the non-inferiority test, and the supremum of the set $\{r \in \mathbb{R} \mid 0 \geq r > E - \delta\}$ for the inferiority test.

| Parasite species | Test | r | WAAVP | Gamma | Binomial | Asymptotic | BNB |
|---|---|---|---|---|---|---|---|
| Hookworm | Inferiority | 0.65 | 0.680 | 0.707 | 0.166 | 0.481 | 0.672 |
| | Non-Inferiority | 0.70 | 0.717 | 0.670 | 0.153 | 0.402 | 0.721 |
| *A. lumbricoides* | Inferiority | 0.90 | 0.416 | 0.489 | 0.062 | 0.324 | 0.440 |
| | Non-Inferiority | 0.95 | 0.242 | 0.177 | 0.019 | 0.083 | 0.342 |
| *T. trichiura* | Inferiority | 0.45 | 0.820 | 0.835 | 0.246 | 0.605 | 0.794 |
| | Non-Inferiority | 0.50 | 0.875 | 0.848 | 0.237 | 0.578 | 0.870 |

All estimated type I and type II error rates for each hypothesis test, statistical method, sample size and simulated parasite are given in Figures S1-S4. The error rates were generally very similar between WAAVP, Gamma and BNB methods for all parasites and sample sizes, except for the simulated efficacy values close to 100%, where the BNB method had a clear advantage. At a sample size of 1,000, all type I error rates at the critical values fell within the range 1.4-3.1% for the BNB, WAAVP and Gamma methods, but for the Asymptotic method these were still around 10%, and for the Binomial method, all exceeded 30%. With a sample size of N = 20, the type I error rates were substantially above 2.5% at the critical values for all methods, except for the WAAVP and BNB methods for non-inferiority tests with hookworm and *T. trichiura*.

As well as focusing on each of the two tests individually, it is also instructive to consider the expected frequency of obtaining each of the four main typologies. These are shown in Figure 2 for a sample size of *N*=91 over a range of relevant simulated efficacy values. As would be expected, typology 1 (reduced efficacy) is most frequently observed at relatively low values of *r*, reducing to a frequency consistent with a notional 2.5% type I error rate at *r* = $T_I$. Conversely, typology 4 (adequate efficacy) is most frequently observed at relatively high values of *r*, reducing to a frequency consistent with a notional 2.5% type I error rate at *r* = $T_A$. Typology 2 (inconclusive) is most frequently observed for values of *r* within or near to the range between $T_I$ and $T_A$. Typology 3 (borderline efficacy) is only rarely observed due to consistently high type II error rates for both individual tests when $T_A \geq \hat{r} > T_I$ (with the exception of the Binomial method, where typology 3 is seen even at relatively small sample sizes due to the spuriously narrow confidence intervals associated with this method [35]). However, typology 3 is observed more frequently than typology 2 for all methods at the higher sample size of 1,000 (data not shown); the larger sample size reducing the width of the derived confidence interval and hence reducing the observed occurrence of the 'inconclusive' typology.

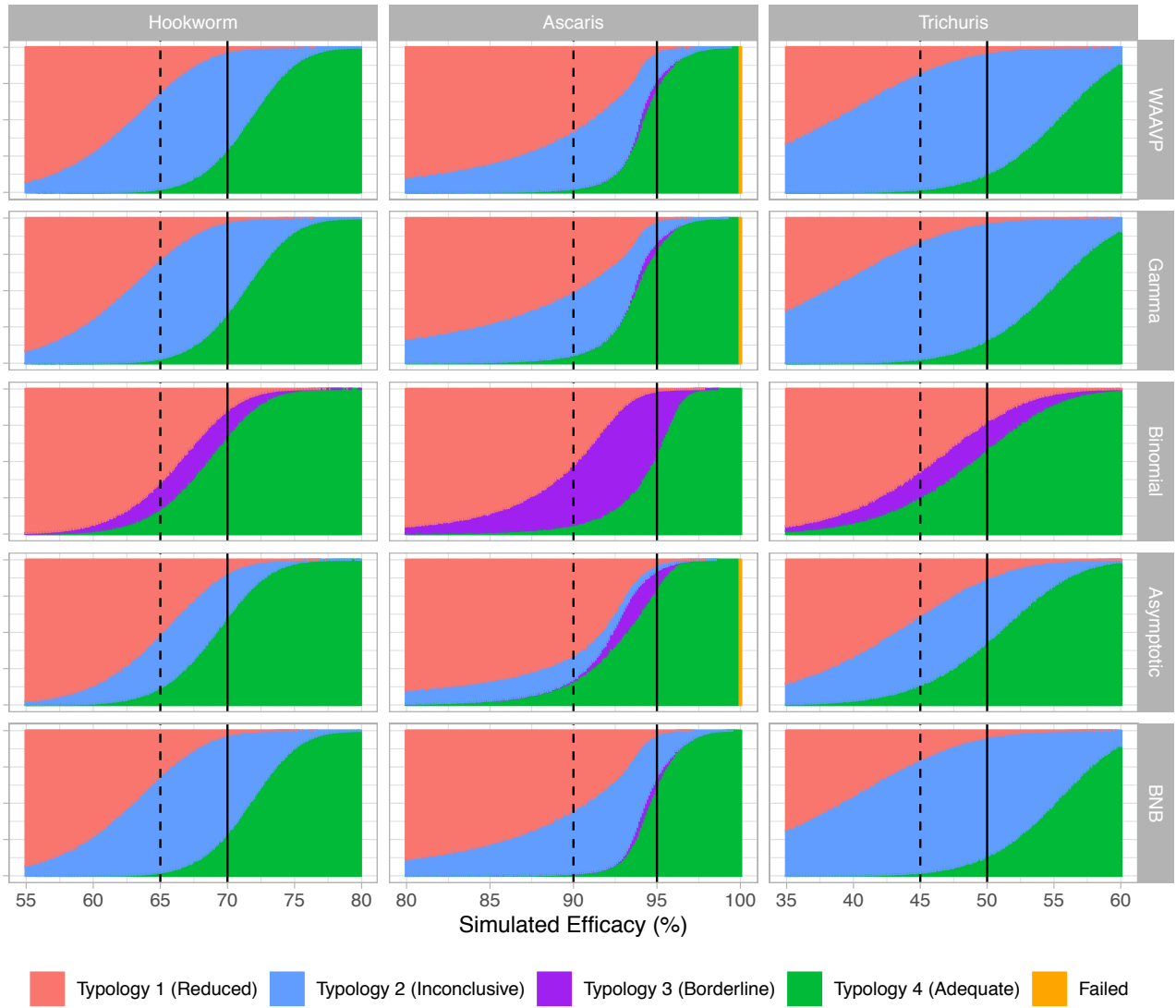

Figure 2: Classification frequencies of the four main typology groups for each of the five statistical methods applied to simulated data with a sample size of N = 91, which corresponds to estimated parameter values from egg reduction rates of three intestinal parasite species in 91 children from Pemba Island (Tanzania). The values corresponding to simulated efficacy values (*r*) equal to $T_A$ and $T_I$ are shown using dashed and solid black vertical lines, respectively, and the x-axis is limited to ± 10 percentage points of these values. The fraction of simulated datasets that could not be analysed using the WAAVP, Gamma and Asymptotic methods due to a 100% observed reduction are shown in orange (a thin vertical stripe close to *r* = 100%, visible only for *Ascaris lumbricoides*).

## 7. Discussion

This work provides a theoretically justified statistical framework within which to evaluate the ratio of means between two data series drawn from negative binomial distributions. Our motivating example is in the field of medical parasitology, and we illustrate our approach by analysing three ERR datasets from a single arm of a randomised controlled trial representing hookworm, *A. lumbricoides* and *T. trichiura* egg counts, with *A. lumbricoides* posing particular statistical challenges due to the observed reduction of 100%. In each case, our framework provides a clearly interpretable and clinically relevant classification for each of the datasets relative to a published target efficacy value. We therefore believe that this framework will be useful within the fields of both medical and veterinary parasitology, as well as other fields where over-dispersed count datasets are commonplace, e.g. occurrence of migraines [3], falls in elderly patients [4], and seizures [5,6].

Although no similar classification framework currently exists within the medical parasitology literature, a classification framework with similar goals forms part of the current WAAVP guidelines for the detection of anthelmintic resistance in animals [20]. It is instructive to briefly compare the typologies within the framework presented here to the WAAVP classification system, which is based on the three possible outcomes of 'resistance present', 'resistance suspected', or 'susceptible'. The key element of the Coles et al. [20] framework is whether the lower 95% confidence limit for the mean is less than a minimum efficacy value [35]. If this is the case, then the treatment will be described as 'resistance present' if $\hat{r} < E$, or 'resistance suspected' if $\hat{r} \geq E$. A lower confidence limit below this minimum efficacy value could reflect resistance, but it could also be driven simply by a small sample size: the strongest interpretation of such an event therefore ought to be that there is no evidence that the treatment is delivering an adequate efficacy. This is a much weaker statement than the recommended classification of the treatment as 'resistance present' or 'resistance suspected', and implies the potential for a great number of false positive conclusions of resistance within the existing literature. As the use of confidence intervals in this framework is not based on a meaningful hypothesis test, the conclusions are not well formulated and the interpretation is unclear. In contrast, one of the important features of the inferiority/non-inferiority framework presented here is the focus on establishing positive evidence for statements made about efficacy, and the explicit acknowledgment of uncertainty, particularly that arising from limited sample sizes. For the interested reader, an interactive tool to explore the consequences of this classification framework is provided at http://www.fecrt.com/framework

Our framework requires two estimates to inform the threshold values used: the target efficacy $E$ and a non-inferiority margin $\delta$. In the context of assessing anthelmintic efficacy, the former should be chosen to reflect the expected ERR associated with a population of parasites that are fully susceptible to the anthelmintic in question: we suggest that this value can be drawn from the literature in most cases. The non-inferiority margin should be chosen as the maximum clinically acceptable difference in efficacy from the target [34], which may vary between different situations and study objectives. In the context of ERR, a value of $\delta = 0.1$ rather than 0.05 is equally justifiable if based on suitable clinical reasoning. Another important aspect of non-inferiority testing as implemented in the medical domain is the choice of confidence level [43]. Whereas traditional inferiority and superiority testing is typically carried out with a 5% significance level (95% confidence), non-inferiority testing is frequently carried out with alternative, lower confidence levels such as 90%. The decision to use smaller confidence limits reduces the threshold of evidence required to reject the null hypothesis and accept the non-inferiority of the treatment; this is particularly important where the margin of equivalence is relatively small. Although ERR results have so far been interpreted exclusively using 95% confidence intervals, there is no reason that this flexibility in choice of confidence level should not be utilised. The inter-relatedness of the simultaneous inferiority and non-inferiority tests as presented here also gives rise to an additional and related point of consideration: for the most relevant observed estimates of $r$, a statistically significant test result (and therefore potential for type I error) can only be observed from one of the two hypothesis tests at a time, i.e. either for the inferiority test when $\hat{r} < T_A$ or the non-inferiority test when $\hat{r} \geq T_I$. If the parameter range $T_A \geq \hat{r} > T_I$ (where there is some possibility of obtaining typology 3) is disregarded, these considerations imply that for the most relevant values of $r$, an alpha value of 5% corresponds to an overall type I error rate of 5%. This choice might therefore be more appropriate than the 2.5% value implied by the current standard use of 95% confidence intervals. Using a higher alpha value would lead to correspondingly higher power, which is of potentially considerable practical benefit. Further work to explore the implications of this would be beneficial: the use of a formal statistical framework such as the one proposed here would be a coherent basis on which to assess such proposals.

In addition to facilitating the interpretation of results following analysis of existing data, the framework developed also allows for the type I and type II error rates of any (computationally simple) statistical method to be quickly and easily quantified. This has been utilised here to show that the Gamma, WAAVP and BNB methods of analysis have broadly equivalent type I and type II error rates for population parameter values relevant to the three datasets presented. However, the extremely high over-dispersion evident in the *A. lumbricoides* dataset results in estimated type I error rates for the inferiority test using each of these methods that substantially exceed the notional 2.5% values. An important exception to these similarities is seen when high simulated efficacy values are used, when the BNB method has a clear advantage due to its ability to meaningfully analyse datasets with a 100% observed reduction. The potential disadvantage of the BNB method is that it does not produce confidence intervals, although this is not necessarily a disadvantage if the aim is to classify the observed reduction in a manner similar to that presented here. The apparently good performance of the WAAVP and Gamma methods is perhaps surprising given that a normal approximation has previously been shown to be poor for egg count data [37], but we note that the relatively large sample size and pre-treatment mean egg counts of our example datasets are likely to favour these methods. We therefore recommend the use of either the Gamma, WAAVP or BNB method for analysing similar datasets, with the latter likely to be preferred for lower pre-treatment mean egg counts. Alternatively, a more computationally-intensive method such as MCMC could be used [23–25,44], although we note the relative difficulty in verifying the properties of statistical tests based on MCMC compared to those based on the computationally simpler methods presented here. For datasets with different expected parameter values, it would seem prudent to undertake a further simulation exercise to verify that the planned statistical method provides valid inference under these conditions: this is likely to be of particular concern for small datasets with relatively high over-dispersion. We also recommend that any further development in statistical methods be verified using the same simulation methods based on parameters informed by similarly appropriate data sets.

Although beyond the scope of this paper, a similar simulation approach could be utilised as an aid to designing a prospective study to help determine the appropriate sample size for a given set of parameter values. This aspect of study design is currently largely neglected within the parasitology community due to the lack of widely available methods, although simulation studies have previously been published for a given set of parameter values [8,21,23,35,45] and methods are available for testing that $r > 0\%$ [7,11,12]. Where each of the two hypothesis tests are considered separately, the results could be interpreted as a standard analysis of statistical power. However, when combining the results of the two related hypothesis tests into a single typology classification, we suggest that something similar to the typology plot given in Figure 2 may be easier to interpret. Comparison of these plots, if calculated for different sample sizes and choices of alpha, would allow practitioners to determine the minimum sample size required for a sufficiently high probability of obtaining a non-inconclusive typology classification over the range of expected efficacy values that are relevant to their application, while controlling the risk of recording a misleading classification. Code to implement these sample size calculations is provided along with implementations of all the statistical methods discussed here as part of the open-source bayescount package for R [42]. In order to facilitate the use of these methods within the wider medical community, user interfaces for both study planning and analysis of existing datasets are provided along with the underlying R package via the following website: http://www.fecrt.com/framework

**Acknowledgements**

The authors are grateful to Professor Sandy Love, Professor Nick Jonsson and the late Professor George Gettinby for discussions that helped to inspire the foundations of the work presented here. We also gratefully acknowledge the assistance of Benjamin Speich in helping to identify an appropriate exemplar dataset, and Sarah Layhe (Vaetta Editing) for proofreading this manuscript. IJM and GTI acknowledge funding from the Scottish Government Rural and Environment Science and Analytical Services Division (RESAS). BL acknowledges post-doctoral fellowship funding from The Research Foundation - Flanders (FWO).

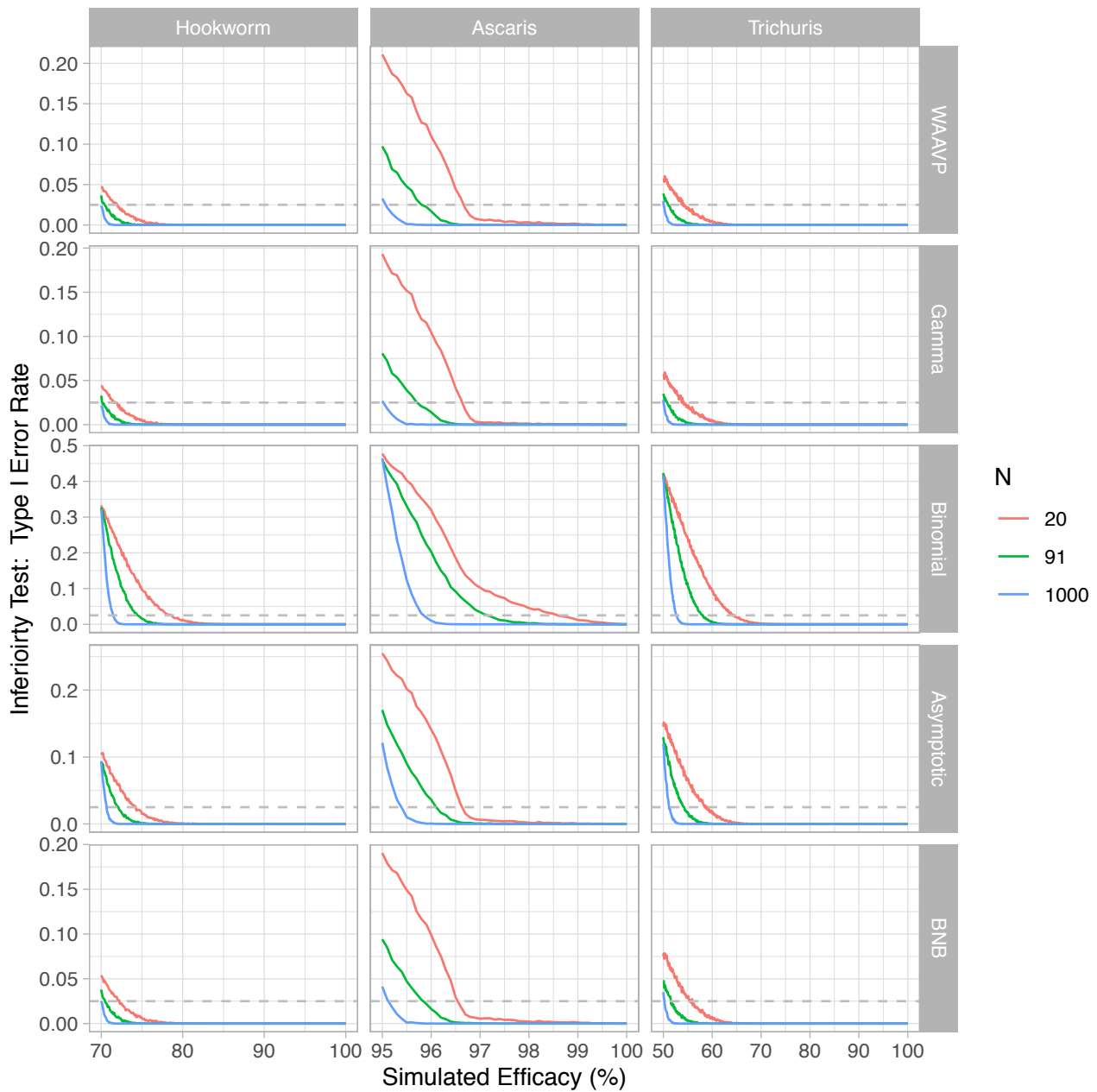

Figure S1: Estimated type I error rates (y-axis) for the inferiority test based on each of the five statistical methods with varying simulated efficacy values (*r*; x-axis) and sample sizes of N = 20, N = 91 and N = 1,000 (colours). Estimates were obtained by Monte Carlo estimation from simulated data corresponding to parameter values from egg reduction rates of three intestinal parasite species in 91 children from Pemba Island (Tanzania). The nominal 2.5% type I error rate applicable at $r = T_I$ (left boundary of the x-axis) is shown as a horizontal dashed line.

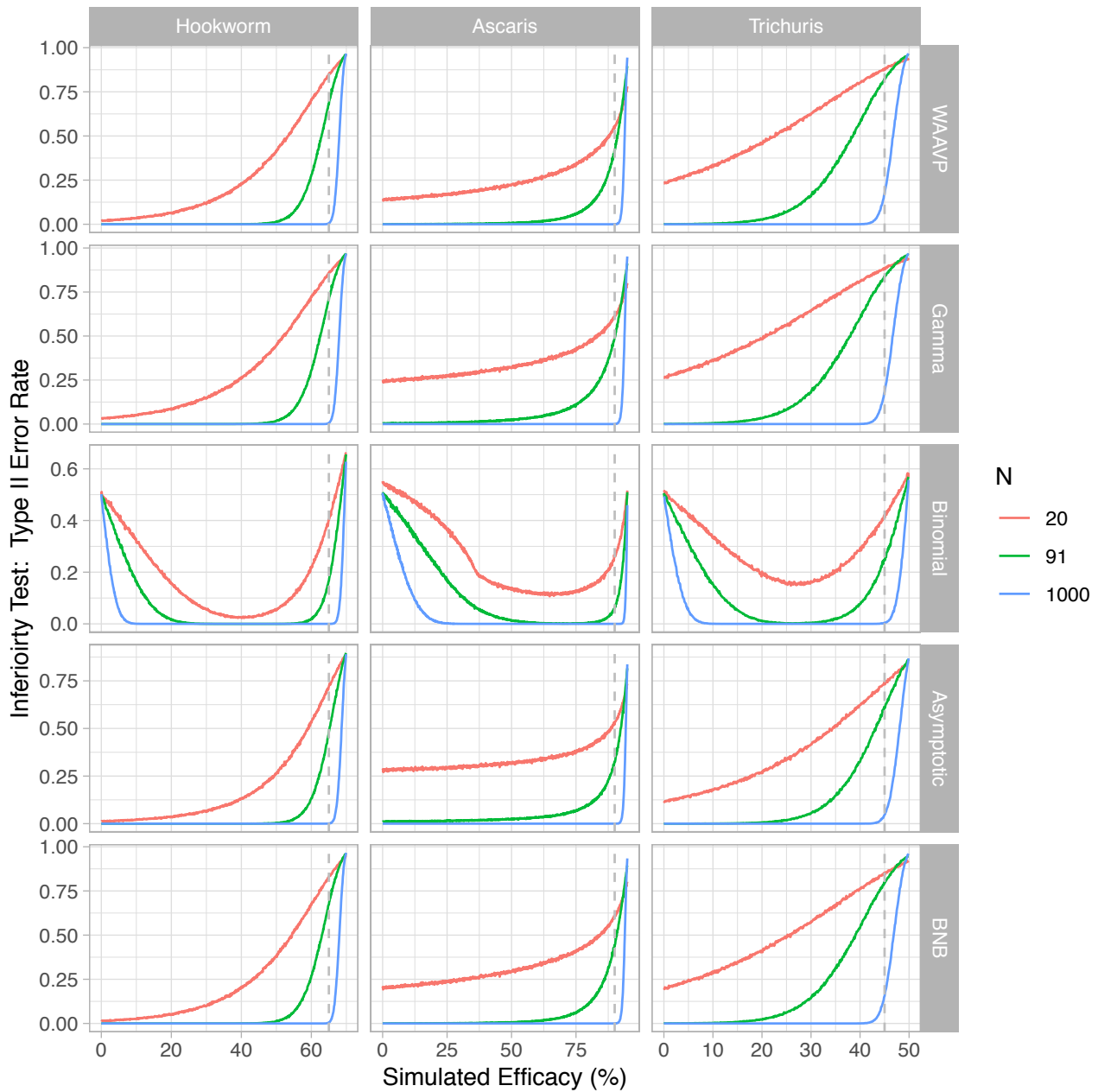

Figure S2: Estimated type II error rates (y-axis) for the inferiority test based on each of the five statistical methods with varying simulated efficacy values (*r*; x-axis) and sample sizes of N = 20, N = 91 and N = 1,000 (colours). Estimates were obtained by Monte Carlo estimation from simulated data corresponding to parameter values from egg reduction rates of three intestinal parasite species in 91 children from Pemba Island (Tanzania). The value of *r* corresponding to *E* is shown as a vertical dashed line.

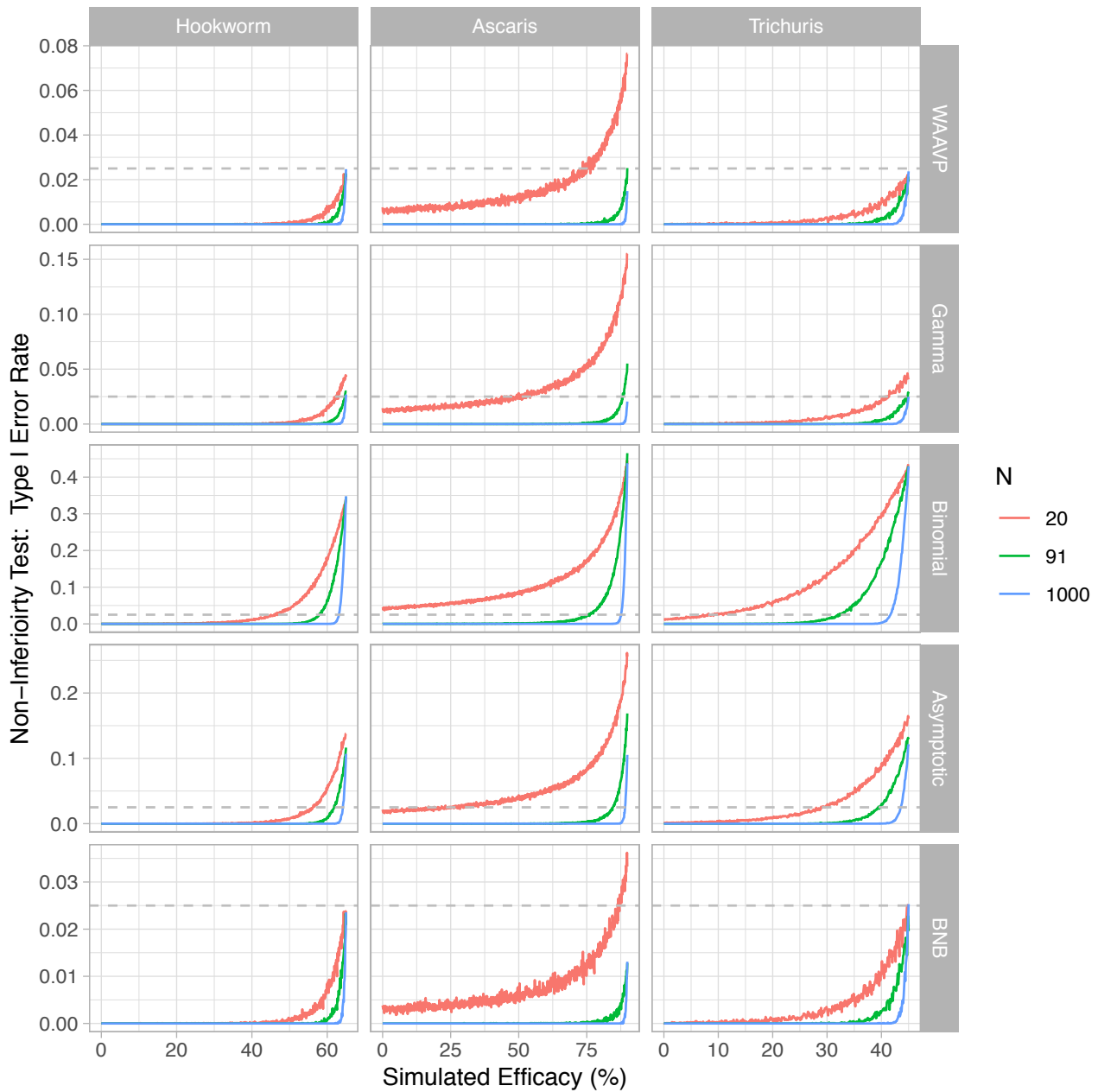

Figure S3: Estimated type I error rates (y-axis) for the non-inferiority test based on each of the five statistical methods with varying simulated efficacy values (*r*; x-axis) and sample sizes of N = 20, N = 91 and N = 1,000 (colours). Estimates were obtained by Monte Carlo estimation from simulated data corresponding to parameter values from egg reduction rates of three intestinal parasite species in 91 children from Pemba Island (Tanzania). The nominal 2.5% type I error rate applicable at $r = T_A$ (right boundary of the x-axis) is shown as a horizontal dashed line.

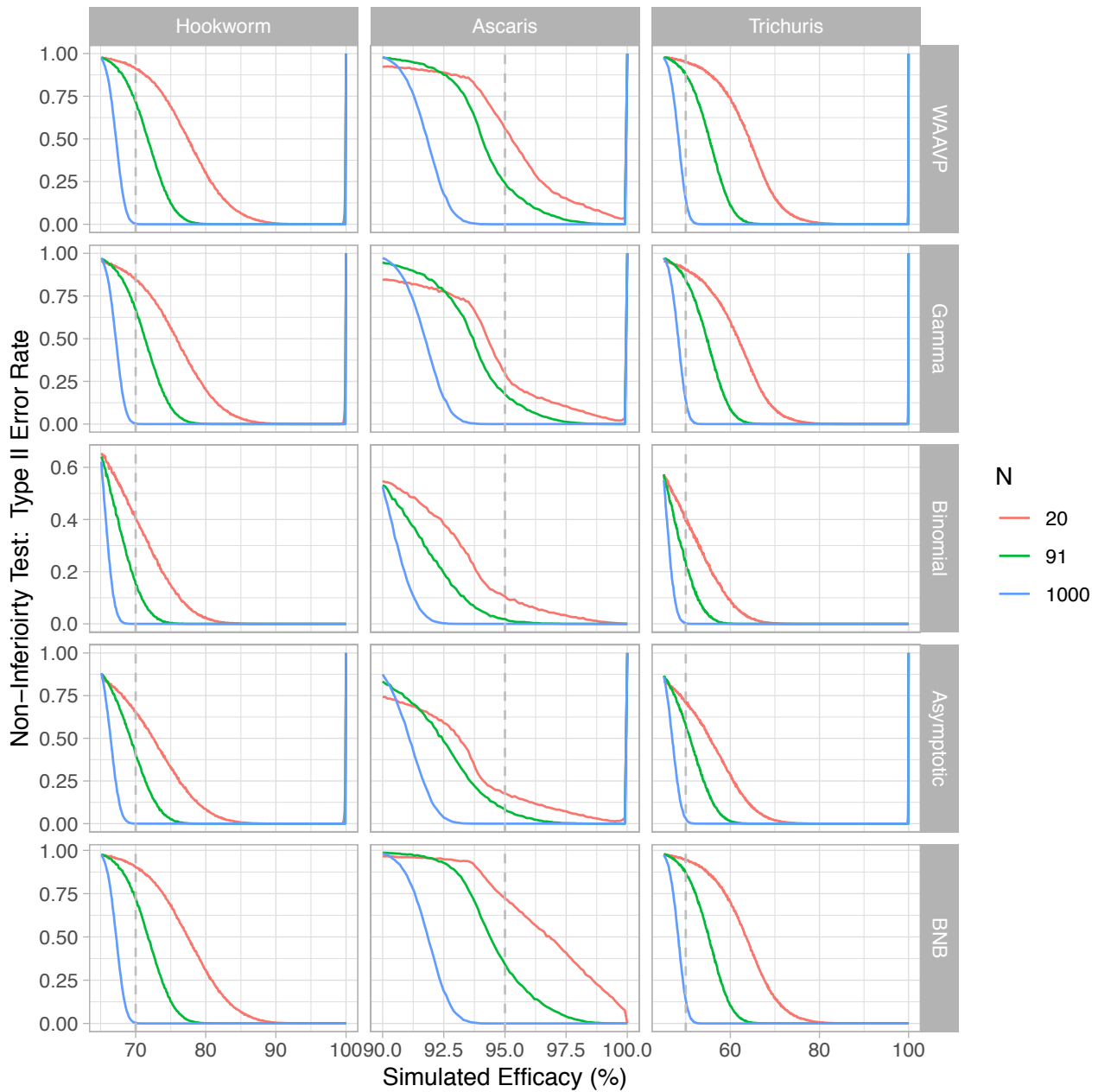

Figure S4: Estimated type II error rates (y-axis) for the non-inferiority test based on each of the five statistical methods with varying simulated efficacy values ($r$; x-axis) and sample sizes of N = 20, N = 91 and N = 1,000 (colours). Estimates were obtained by Monte Carlo estimation from simulated data corresponding to parameter values from egg reduction rates of three intestinal parasite species in 91 children from Pemba Island (Tanzania). The value of $r$ corresponding to $E - \delta$ is shown as a vertical dashed line.

## Statistical Appendix

This appendix contains the derivation of the beta negative binomial (BNB) method presented in "*A hypothesis testing framework for the ratio of means of two negative binomial distributions: classifying the efficacy of anthelmintic treatment against intestinal parasites*" by Denwood et al.

Background

The efficacy $r$, defined as the ratio of the means of the two count distributions $x_i$ and $y_i$, is typically the statistic of interest. However, within a hypothesis testing framework, we may also usefully consider the distribution of a sufficient statistic for the mean of one of the count distributions integrated over all possible values of an equivalent statistic for the other count distribution. For this derivation, we assume that both count distributions $x_i$ and $y_i$ follow negative binomial distributions, and that the over-dispersion parameters $k_1$ and $k_2$ are either known or can be estimated from the data. We show here how a beta negative binomial compound distribution can be used to describe the distribution of $y_i$ under one or more null hypotheses, with parameters that can be derived from the observed data $x_i$ and given parameter values $r$, $k_1$ and $k_2$.

Estimation of the expectation for $\mu_2$

We use the formulation of the negative binomial distribution defined as the number of successes before a given number of failures, with probability of success $p$, which is generalised to allow any strictly positive value of number of failures. This has the same distribution function as a gamma-Poisson distribution, with shape parameter $k$ representing the number of failures, and mean number of successes given by $\mu = \frac{pk}{1-p}$. We assume that $k_1$ and $k_2$ are known, while we treat $p_1$ as a random variable with properties depending on the observed $x_i$. In the following, we use a Bayesian framework to fully describe the posterior distribution of $p_1$ using a conjugate prior distribution. Specifically, we use a beta distribution as the conjugate prior for the negative binomial (given fixed $k$), with posterior parameters:

$$p_1 \sim \text{Beta}(\alpha_1, \beta_1) \qquad \alpha_1 = \alpha_0 + \sum_{i=1}^{n} x_i \qquad \beta_1 = \beta_0 + kn$$

where $\alpha_0$ and $\beta_0$ are the hyperparameters for the prior for parameter $p_1$. This line of reasoning follows that of Dobson et al. [1], who use a beta distribution to derive the confidence interval for $r$ based on the assumption that

$$\sum_{i=1}^{n} y_i \sim \text{Binomial}(\sum_{i=1}^{n} x_i, 1-r).$$

*Hypothesis testing for the ratio of negative binomial distributions: statistical appendix*

We can then derive the uncertainty distribution for the post-treatment probability of success $p_2$ by treating it as a non-linear function of the beta-distributed pre-treatment probability of success $p_1$. Expressing $\mu_2$ as a function of $p_1$, $k_1$, $k_2$ and $r$, it can be seen that:

$$\mu_2 = \frac{p_2 k_2}{1 - p_2} = (1-r)\mu_1 = (1-r)\frac{p_1 k_1}{1 - p_1}.$$

Rearranging yields $p_2$ as a non-linear function $g$ of $p_1$, $r$, $k_1$ and $k_2$:

$$p_2 = g(p_1; k_1, k_2, r) = \frac{p_1 k_1 (1-r)}{p_1 k_1 (1-r) - p_1 k_2 + k_2}.$$

There are two options to yield the full distribution of the uncertainty in $p_2$ given the beta conjugate prior for $p_1$ (and fixed $k_1$, $k_2$, $r$):

1. Numerical integration based on a sample of realisations of $p_1$ from Beta($\alpha_1$, $\beta_1$) with subsequent functional transformation, which yields a Monte Carlo approximation to the distribution of $p_2$.
2. Using an approximation based on the delta method to approximate the mean and variance of the non-linear, smooth transformation $p_2 = g(p_1; k_1, k_2, r)$, given the fixed values and the moments of the random variable $p_1$.

Given that the primary objective here is to provide a method of performing power calculations, we prefer the computationally simpler method (2), and give the derivation of this below. However, we acknowledge the utility of numerical integration methods, and have also used these to verify that the approximations required in the delta method are justified.

Delta method approximation

The mean of the distribution of $g(p_1; k_1, k_2, r)$ conditional on the Beta($\alpha_1$, $\beta_1$) distribution of the random variable $p_1$, is empirically well estimated by the first 2 terms of the Taylor series expansion:

$$E(p_2) \approx g(E(p_1)) + \frac{1}{2}g''(E(p_1)) \cdot E\left((p_1 - E(p_1))^2\right).$$

The variance of $g(p_1; k_1, k_2, r)$ is less stable in estimation, but is empirically well approximated using the higher order Taylor series expansion given on page 11 of Cooch & White [2] as:

$$Var(p_2) \approx g'(E(p_1))^2 \cdot E\left((p_1 - E(p_1))^2\right)$$

$$+ 2g'(E(p_1)) \cdot \frac{g''(E(p_1))}{2} \cdot E\left((p_1 - E(p_1))^3\right)$$

$$+ \left(\frac{g''(E(p_1))^2}{4} + 2g'(E(p_1)) \cdot \frac{g'''(E(p_1))}{3!}\right) \cdot E\left((p_1 - E(p_1))^4\right)$$

$$+ \left(2g'(E(p_1)) \cdot \frac{g''''(E(p_1))}{4!} + g''(E(p_1)) \cdot \frac{g'''(E(p_1))}{3!}\right) \cdot E\left((p_1 - E(p_1))^5\right).$$

The required 1st-4th partial derivatives of $g$ with respect to $p_1$ given fixed and strictly positive $k_1$, $k_2$, $r$ can readily be derived as:

$$\frac{\partial g}{\partial p_1} = \frac{k_1 k_2 (1-r)}{\left((p_1-1)k_2 - p_1 k_1(1-r)\right)^2}$$

$$\frac{\partial^2 g}{\partial p_1^2} = \frac{-2 k_1 k_2 (1-r)(k_2 - k_1(1-r))}{\left((p_1-1)k_2 - p_1 k_1(1-r)\right)^3}$$

$$\frac{\partial^3 g}{\partial p_1^3} = \frac{6 k_1 k_2 (1-r)(k_2 - k_1(1-r))^2}{(p_1 k_1(1-r) - p_1 k_2 + k_2)^4}$$

$$\frac{\partial^4 g}{\partial p_1^4} = \frac{24 k_1 k_2 (1-r)(k_1(1-r) - k_2)^3}{\left((p_1-1)k_2 - p_1 k_1(1-r)\right)^5}$$

The 2$^{nd}$, 3$^{rd}$, 4$^{th}$ and 5$^{th}$ central moments of $p_1$ are required to estimate the variance of $g(p_1; k_1, k_2, r)$. These can be reformulated in terms of crude moments by expanding the binomial; for example, for a random variable $X$ with mean $\mu$:

$$E((X-\mu)^3) = E(X^3) - 3\mu E(X^2) + 2\mu^3.$$

Given that $p_1$ is Beta($\alpha_1$, $\beta_1$) distributed, we have:

$$E(p_1^v) = \frac{\Gamma(\alpha_1 + \beta_1)\,\Gamma(\alpha_1 + v)}{\Gamma(\alpha_1)\,\Gamma(\alpha_1 + \beta_1 + v)},$$

which for integer $v$ has a trivial closed form; for example, $v = 3$ gives:

$$E(p_1^3) = \frac{\Gamma(\alpha_1+\beta_1)\,\Gamma(\alpha_1+3)}{\Gamma(\alpha_1)\,\Gamma(\alpha_1+\beta_1+3)} = \frac{(\alpha_1+2)(\alpha_1+1)\alpha_1}{(\alpha_1+\beta_1+2)(\alpha_1+\beta_1+1)(\alpha_1+\beta_1)}.$$

All of the required quantities can therefore be easily calculated so that the mean and variance of the uncertainty distribution describing $p_2$ (and therefore also that for $\sum_{i=1}^{N_2} y_i$) can be derived. Although we have not defined the functional form of this distribution of uncertainty for $p_2$, the distribution is bounded at [0, 1] and numerical integration of the transformation empirically approximates a beta distribution within sensible parameter bounds. We therefore describe the distribution of uncertainty for $p_2$ using a Beta($\alpha_2$, $\beta_2$) distribution with parameters derived from the mean and variance of this distribution as defined above.

Distribution of the test statistic

In order to implement the statistical tests, a sufficient statistic derived from $y_i$ must be compared to the distribution of the same statistic under the null hypothesis. We note that the sum of $y_i$ is distributed according to:

$$\sum_{i=1}^{N_2} y_i \sim \text{NegBin}(N_2 k_2, p_2).$$

The quantity $N_2k_2$ is assumed to be fixed, and $p_2$ is described by a Beta($\alpha_2, \beta_2$) distribution, conditional on the observed $x_i$, fixed quantities $k_1$ and $k_2$, prior parameters $\alpha_0$ and $\beta_0$, and an assumed fixed value for the efficacy $r$ that corresponds to the efficacy specified under the null hypothesis. We can therefore describe the expected distribution of the test statistic using a beta negative binomial distribution, defined as $s$ successes before a given number of failures for consistency with our use of the negative binomial distribution, with probability mass function given by:

$$p(s; \alpha_2, \beta_2, N_2k_2) = \frac{\Gamma(N_2k_2 + s)}{k!\,\Gamma(N_2k_2)} \frac{B(\beta_2 + N_2k_2, \alpha_2 + s)}{B(\beta_2, \alpha_2)},$$

where $B$ is the beta function. The observed test statistic $\sum_{i=1}^{N_2} y_i$ can be evaluated with respect to this beta negative binomial distribution to derive the associated p-value.